# Modifying the surface electronic properties of YBa$_2$Cu$_3$O$_{7-\delta}$ with cryogenic scanning probe microscopy

S. Urazhdin[1], W. K. Neils[2], S. H. Tessmer[1], Norman O. Birge[1] and D. J. Van Harlingen[2]

[1]Department of Physics & Astronomy and Center for Fundamental Materials Research, Michigan State University, East Lansing, MI 48824
[2]Department of Physics, University of Illinois at Urbana-Champaign, Urbana, IL 61801

**Abstract**
We report the results of a cryogenic study of the modification of YBa$_2$Cu$_3$O$_{7-\delta}$ surface electronic properties with the probe of a scanning tunneling microscope (STM). A negative voltage applied to the sample during STM tunneling is found to modify locally the conductance of the native degraded surface layer. When the degraded layer is removed by etching, the effect disappears. An additional surface effect is identified using Scanning Kelvin Probe Microscopy in combination with STM. We observe reversible surface charging for both etched and unetched samples, indicating the presence of a defect layer even on a surface never exposed to air.

**PACS classification codes:** 68.37.Ef; 74.76.Bz

## 1. Introduction

Soon after the discovery of materials exhibiting High Temperature Superconductivity (HTSC), it was found that changes in the oxygen content from exposure to ambient air resulted in a degraded surface layer [1]. Many subsequent studies have investigated the effects of the electromagnetic environment on the surface both to elucidate the nature of the superconductivity and to test the stability of the materials for use in applications. A number of effects have been observed including photodoping resulting in persistent photoconductivity [2], the photovoltaic effect [3], and electric-field-induced conductance changes [4, 5]. Although a clear relationship between these properties and the phenomenon of HTSC has yet to emerge, it has been firmly established that oxygen concentration and disorder play a crucial role in determining the electronic properties. According to one of the proposed mechanisms [6], light or electric field directly interacts with the charge carriers, changing their concentration. Electromigration of oxygen ions represents another possibility. Such a rearrangement is shown to be possible due to a small oxygen migration energy and high density of vacancies in the Cu-O chains [7].

Several studies have investigated a number of surface modification effects that can be induced in a HTSC material using a scanning probe [8]. In contrast to standard transport techniques, the scanning probe methods allow the modification to be imaged directly. Here we report cryogenic Scanning Tunneling Microscopy (STM) and Scanning Kelvin Probe Microscopy (SKM) observations of modifications of the YBa$_2$Cu$_3$O$_{7-\delta}$ surface electronic properties induced by the STM tip. To the best of our knowledge, the Kelvin probe component of this work – a measure of surface charge distributions – represents the first cryogenic SKM study of a HTSC material. We find that a negative voltage applied to the sample during STM tunneling locally alters the insulating properties of the native degraded surface layer, consistent with recent transport studies. In addition, SKM measurements show reversible surface charging, an effect that persists even with the degraded layer removed.

## 2. Overview

We have performed a series of STM and SKM measurements on YBa$_2$Cu$_3$O$_{7-\delta}$ (001) films grown by laser ablation on SrTiO$_3$ substrates. All the data reported here were acquired on a 100 nm thick underdoped sample with a transition temperature of 78 K. Similar results, however, were obtained on an optimally-doped sample with a transition temperature near 90 K.

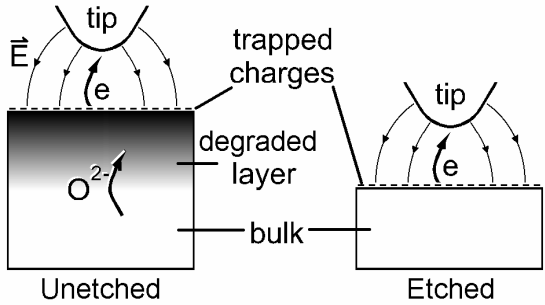

**Figure 1**. Schematic of the two surface types studied. "Unetched" samples were subjected to dry air for several days and contain a degraded surface layer. "Etched" samples had 15 nm of the surface layer removed prior to the measurements. The current and electric field corresponding to the STM tunneling process is also shown schematically.

To distinguish the effects of the degraded surface layer from bulk behavior, we prepared the surfaces in two different ways, shown schematically in Fig. 1. "Unetched" samples were stored in dry air for several days after the ablation, then cleaned with acetone and isopropyl alcohol, and dried in nitrogen gas prior to transfer to the microscopy system. For "etched" samples, the exposed surface layer was removed by etching the samples for 15 seconds (at a rate of 1 nm/s) in 1 % Br solution in absolute methanol. Using a glove box attached to the system, the samples were then rinsed in absolute methanol, dried in ultrapure He gas, and then directly transferred to the microscope without any exposure to laboratory air. This Br etching technique has been used to prepare HTSC surfaces for low-resistance ohmic contacts [9] as well as STM spectroscopic measurements [10].

### 3. Scanning probe methods

A custom-built low-temperature STM with integrated sample processing was employed for the measurements [11]. All measurements were performed with the sample and tip immersed directly in liquid He-4 at 4.2K. This environment provides for excellent temperature stability and minimizes the possibility of surface contamination during the experiment. In addition to standard STM spectroscopic measurements, we have also performed effective work function $\phi_{eff}$ measurements by monitoring the tunneling current as a function of tip height above the surface [12]. We acquire these data by modulating the tip height at a frequency above the STM feedback bandwidth, and then recording the resulting tunneling current modulation as a function of the tip position. This measurement provides complimentary information to the spectroscopic data. For example, in addition to the vacuum barrier, electrons will tunnel through a sufficiently thin layer of perfect insulator on top of a metallic sample; hence spectroscopic measurements alone will not indicate the presence of the insulating surface. However, the effective work function measurement is sensitive to tip-insulator mechanical interactions, which result in a reduction of the measured $\phi_{eff}$ compared to the typical metallic value of several eV. The presence of such an insulating layer represents a major concern in the interpretation of the STM measurements of the HTS surfaces [13].

Lastly, we performed surface potential measurements employing Scanning Kelvin Probe Microscopy [14]. A low-temperature charge sensor circuit with a sensitivity of about 0.03 *electrons*/√Hz [15] was used for this measurement. The tip was positioned out of tunneling range, typically about 5 nm above the surface, and its height was modulated with an rms amplitude of about 1 nm. The tip-surface capacitive coupling resulted in ac current. The local surface potential value was then determined by minimizing the current as a function of bias voltage. The measured surface potential $V_s$ can be understood as the sum of two contributions: (1) the work function difference between the tip and sample (contact potential), and (2) the potential from trapped surface charges [16].

### 4. Results

Fig. 2(a) shows a map of the relative differential conductance over a 3 μm x 3 μm area of an unetched sample; brighter shades indicate greater dI/dV. The image was obtained by monitoring dI/dV while scanning the tip with the current maintained at I= 2 pA and a bias voltage V =-5 V applied to the sample. A clear low-conductance (dark) feature appears in the center. This feature corresponds to a 0.75 μm x 0.75 μm square that had been modified by excessive STM tunneling prior to acquiring the image. To quantify the effect in detail, Fig. 2(b) shows two representative differential conductance curves as a function of V. For unmodified areas, dI/dV is relatively small at low voltages and increases considerably for |V|>1 V. The result is a quadratic U-shaped function shown by the squares. In contrast, excessive tunneling alters the shape of the spectra, as shown by the dots. In this case, prior to acquiring the spectra, the sample was modified by seven separate scans of STM tunneling over a 0.3 μm x 0.3 μm area with I = 20 pA, V= -3 V. For STM spectra, the magnitude of the curves depends on the current and voltage set points, which fix the nominal tunneling resistance prior to acquiring each data set. By integrating the unmodified and modified curves of Fig. 2(b), we normalized each to represent the same current at –0.5 V. We see that the modified area yields more linear spectra in the voltage range of ±1.25 V, resulting in more V-shaped curves. This behavior is similar to non-locally resolved observations reported by Plecenik and coworkers using $YBa_2Cu_3O_{7-\delta}$ /Au planar tunnel junctions and point contacts [5].

We emphasize that we only observe the modification of dI/dV for negative applied sample voltage, and for unetched samples; differential conductance and effective work function measurements of etched samples show no significant changes with STM scanning. Clearly the effect is connected to properties of the surface degraded layer present for unetched samples. For data acquired in the voltage range ~ ±1 V, the superconducting gap is not discernible. However, we do observe gap-like spectra when probing energies on the mV scale, for which we bring the tip closer to the sample to decrease the junction resistance, as shown in the inset of Fig 2(b). Lastly, we note that concurrent measurements of the differential conductance and the effective work function $\phi_{eff}$ show that the decrease in dI/dV is accompanied by an increase in $\phi_{eff}$.

Turning now to the Kelvin probe measurements, Fig. 3 shows two images of the surface potential acquired using the SKM technique. Fig. 3(a) is a grayscale plot of $V_s$ acquired over an unetched sample, with the sample bias adjusted to be close to the contact potential. Fig. 3(b) shows a $V_s$ image of the same area after the surface was altered by STM tunneling into a small 0.3 μm x 0.3 μm area in the center with a voltage of V=-3 V and current of 20 pA (one scan). The altered feature of $\Delta V_s \approx 0.40$ V appears as a dark spot in the center. We find that the surface potential can be reproducibly modified in this way. In general, by tunneling again with inverted sample bias voltage we are able to restore the surface potential to the original value and thus return to the original uniform map. However, starting from an unmodified surface, a positive sample bias voltage does not produce a significant increase in surface potential. Lastly, in contrast to the dI/dV - $\phi_{eff}$ modification described above, we find that this effect occurs for both etched and unetched samples.

To further elucidate the two effects, we study their evolution as a function of scan number, i.e., the number of times the tunneling tip is rastered across a given area. Fig. 4(a) shows the effective work function for etched and unetched samples as a function of the number of times the tip was rastered across a 0.3 μm x 0.3 μm area at a speed of 0.3 μm/s. For each scan, the tunneling conditions were I=20 pA, V= –3 V. We see that for the unetched sample, $\phi_{eff}$ increases for each scan, until the effect approaches saturation at around six scans. The decrease in concurrently acquired dI/dV measurements (not shown) follow a similar pattern, also saturating near six scans. The effective work function measurements for the etched sample start out at a high value and only slightly decrease with scan number. We find that the number of scans over which the differential conductance and effective work function saturates scales roughly proportionally to the speed of the tip as it scans along the surface. In other words, as expected intuitively, the key parameter is the time for which the surface is locally exposed to STM tunneling.

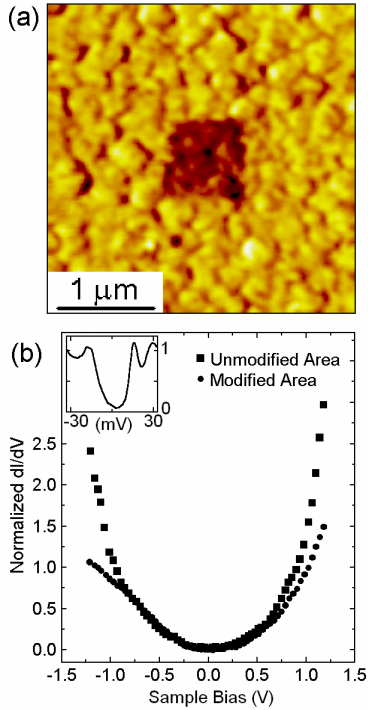

**Figure 2**. Modification of the differential conductance of an unetched sample. (a) A dI/dV map acquired at I=2 pA, V= –5 V over an area for which a central square of 0.75 μm x 0.75 μm had been modified with excessive STM tunneling. Brighter shades indicate greater differential conductance. The modification clearly resulted in a local decrease in dI/dV. (b) An average of 1024 differential conductance spectra acquired over a 0.3 μm x 0.3 μm area of the sample, at a different location than (a). The squares represent spectra acquired prior to modifying the area; the dots represent spectra acquired after modification. The two curves were normalized to correspond to the same current at V=–0.5 V. The modification consisted of scanning over the area with the tunneling current and bias voltage set at I=20 pA and V= -3 V, respectively. The tip was rastered over the area seven times moving at a speed of 0.12 μm/s for each of 128 lines. (inset) Example of gap-like spectra observed when probing energies on the mV scale.

Fig. 4(b) shows the change of the surface potential $V_s$ with scan number, measured with the SKM technique. For samples of both types, it decreases significantly over the first two scans and then remains roughly constant. It is apparent that the modification time of the $V_s$ change is much shorter than that for the dI/dV - $\phi_{eff}$ changes shown in (a). This provides further evidence that the two effects we observe reflect distinct modifications of the sample.

The experimental findings can thus be summarized as follows: STM tunneling with a negative sample voltage induces a concurrent modification of dI/dV and $\phi_{eff}$. The changes are only observed for unetched samples and cannot be reversed by applying a positive sample bias. STM tunneling with negative bias also induces a modification of $V_s$. In contrast, this change is seen in both etched and unetched sample; it occurs on a shorter time scale; and it can be removed by applying a positive voltage.

## 5. Discussion

### 5.1. Differential conductance modification

The fact that the dI/dV - $\phi_{eff}$ effect shown in Fig. 2 and Fig. 4(a) only occurs for unetched samples indicates that it is a property of the degraded surface layer. A satisfactory explanation must account for this and also explain the connection between the differential conductance and the effective work function. We attribute the behavior to the interplay between scattering states in the degraded surface layer and electrostatic bending of the tunneling barrier. We believe the tip is actually in contact with an insulating surface layer while probing the unmodified areas. Surprisingly, relative to the modified areas, the resulting reduced tunneling barrier leads to an *increase* in the measured dI/dV at high bias, as discussed below.

Our interpretation is based on spectra typically observed for planar YBCO tunnel junctions, which exhibit a V-shape at low voltages – likely caused by a broad energy distribution of inelastic scattering states in the barrier [17]. In contrast, we observed U-shaped curves in unmodified areas, such as the squares in Fig. 2(b). This indicates that an additional factor is at play in our STM tunnel junction. A low value of the surface work function, which defines the tunneling barrier, can lead to such effects. For typical STM spectroscopy measurements, as a result of the electrostatic bending of the vacuum barrier, the differential conductance increases at bias voltages comparable to the work function.

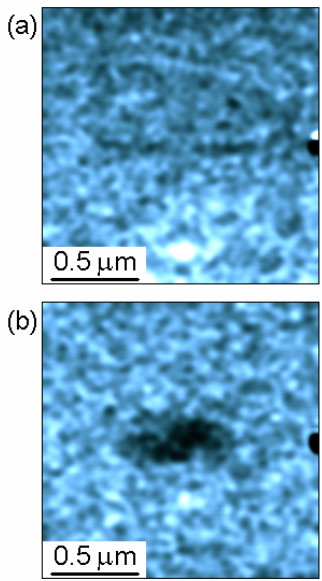

**Figure 3**. SKM map of $V_s$ for an unetched sample, with the bias voltage adjusted to null the contact potential. Brighter shades indicate greater $V_s$. Image (a) shows the unmodified area. Image (b) shows $V_s$ after tunneling over a 0.3 μm x 0.3 μm area in the center with a negative sample bias of V= -3 V and a current of I=20 pA . The contrast between the dark area in the center and the background is about 0.4 V.

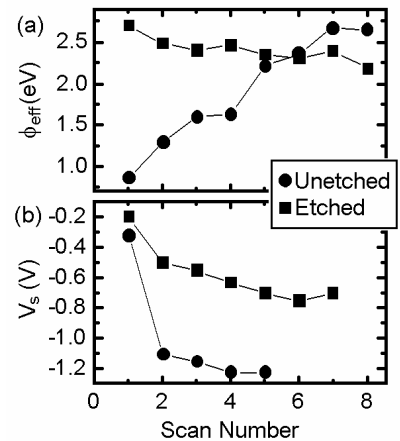

**Figure 4**. Modification characteristics for both unetched and etched samples. The effective work function $\phi_{eff}$ (a) and surface potential $V_s$ (b), averaged over an area of 0.3 μm x 0.3 μm, are plotted as a function of the number of times the tip rastered over the area while tunneling. The STM tunneling conditions were I=20 pA, V= –3 V, and the scan speed was 0.3 μm/s for each of 128 lines.

Kelvin probe measurements performed on our samples, however, give the work function value of about 5 eV, assuming the work function of the PtIr tip of about 5.6 eV [18]. Hence, a U-shaped dI/dV spectrum would only be expected if the bias voltage range were on the scale of a few volts. However, if the tip is in contact with the sample, the effective energy barrier will be reduced, yielding a U-shaped spectrum at lower voltage ranges. We believe that this is the case for our STM measurements in unmodified areas. We therefore interpret the surface modification due to excessive STM tunneling at negative bias as an increase in the conductivity of the degraded layer. This gives an increased tip-sample separation so that vacuum tunneling is achieved – explaining both the more V-shaped spectra and the measured increase in $\phi_{eff}$ in modified areas.

Two possible mechanisms may cause the enhanced conductivity of the surface layer: the tip's applied electric field or the injected tunneling current, as schematically shown in Fig. 1. Based on the sensitivity to the polarity of the applied voltage, we believe the effect originates from electric field. Following the conclusions of the study by Plecenik and coworkers [5], we suggest that the tip's electric field ~$5 \times 10^7$ V/cm may cause oxygen electromigration toward the surface from underlying layers. The resulting oxygen replacement in a deficient surface layer would tend to restore its electronic properties. This scenario is indicated in Fig. 1.

Lastly, returning to Fig. 4(a), we note that the above explanation does not address the decrease in $\phi_{eff}$ for etched samples. This behavior is consistent with an actual work function change as discussed below.

*5.2. Surface potential modification*

To account for the reduction of surface potential shown in Fig. 3 and Fig. 4(b), we begin by noting that the tip is positioned well out of tunneling range for the SKM measurements. Hence the method is only sensitive to electric field, and no other tip-sample interactions are expected to contribute. This leads to a straightforward interpretation of the $V_s$ reduction as the result of trapped charges near the surface – similar to observations in semiconducting samples [16]. Again based on the sensitivity to the polarity of the applied voltage, we believe the mechanism for the effect is the electric field. For negative sample bias voltage, the field attracts electrons to the surface, some of which become trapped in impurity or defect states. Because the effect is more pronounced in unetched samples, we speculate that the trapping states originate from oxygen disorder at the exposed surface. We can estimate the density of the trapped charges from the magnitude of the potential decrease $\Delta V_s$, as described below.

Charges trapped at the exposed surface will produce an electric field that will be screened by the underlying metallic bulk, creating a dipole layer. This will contribute to the surface potential measured by the Kelvin probe as $\Delta V_s = enr/\varepsilon_0$, where $n$ is the surface density of the charges, and $r$ is the characteristic screening radius of the HTSC material. Using a screening radius of $r \approx 1$ nm [19], and $\Delta V_s \approx 0.5$ V, corresponding to the measured magnitude for etched samples, we find $en \approx 4 \times 10^{-7}$ C/cm$^2$. This corresponds to an average separation of trapped charges of about 7 nm.

## 6. Conclusion

We have performed a series of low-temperature scanning probe microscopy measurements to characterize the surface properties of YBa$_2$Cu$_3$O$_{7-\delta}$ thin films. Spectroscopic and effective work function measurements of air-exposed samples indicate the presence of an insulating degraded surface layer. We find that repetitive STM tunneling can modify the electronic properties of this layer. In particular, the application of negative sample bias voltage during tunneling results in an increase of the surface conductivity. We believe that this effect may be caused by the electromigration of oxygen atoms from the bulk of the film toward the surface, partially restoring an oxygen deficiency in the surface layer. An additional surface effect has been identified even for samples that have 15 nm of the surface removed: Using SKM in combination with STM, we observe reversible surface charge trapping. This observation implies the presence of a defect layer susceptible to charging by the high electric field of the STM probe.

**Acknowledgments**

We thank L. H. Greene, J. Mannhart, and J. Nogami for helpful discussions. This work was supported by the National Science Foundation grants DMR99-72087 and DMR00-75230 (in part). SHT acknowledges support of the Alfred P. Sloan Foundation.